\documentstyle[aps,pra,epsf]{revtex}

\begin{document}
\twocolumn[\hsize\textwidth\columnwidth\hsize\csname %
@twocolumnfalse\endcsname
\draft
\widetext 
\title{Current driven switching of magnetic layers} 

\author{C. Heide}

\address{Department of Physics, New York University, 4 Washington Place, 
New York, NY 10003 } 

\author{P. E. Zilberman}

\address{Institute of Radio-Engineering \& Electronics, 
Russian Academy of Sciences,\\
Fryazino, Vvedenskii Sq.\ 1, Moscow Region 141120, Russia } 

\author{R. J. Elliott}

\address{University of Oxford, Department of Physics,\\
Theoretical Physics, 1 Keble Road, Oxford OX1 3NP, United Kingdom}
\date{\today}

\maketitle

\begin{abstract}

The switching of magnetic layers is studied under the action of a spin
current in a ferromagnetic metal/non-magnetic metal/ferromagnetic
metal spin valve. We find that the main contribution to the switching
comes from the non-equilibrium exchange interaction between the
ferromagnetic layers. This interaction defines the magnetic
configuration of the layers with minimum energy and establishes the
threshold for a critical switching current.  Depending on the
direction of the critical current, the interaction changes sign and a
given magnetic configuration becomes unstable. To model the time
dependence of the switching process, we derive a set of coupled
Landau-Lifshitz equations for the ferromagnetic layers. Higher order
terms in the non-equilibrium exchange coupling allow the system to
evolve to its steady-state configuration.

\end{abstract}

\phantom{.}
]
\narrowtext

\section{Introduction}
The possibility of using the exchange field of a spin-polarized
current to aid in switching of a magnetic layer is not only an
intruiging prospect for future applications in the field of
magneto-electronics but also a challenge with interesting physics.
There is still a lot of ambiguity in the interpretation of the complex
data of the recently observed switching of domains in spin
valves~\cite{bi253} and tunnel junctions~\cite{bi235}, of magnetic
clusters~\cite{bi262} and layers~\cite{bi267}, nonetheless, in
Refs.~\cite{bi253,bi262,bi267} it has been argued that the switching
occurs through relaxation of conduction electron spin-polarization to
the local moments of the ferromagnetic layers as proposed by
Slonczewski~\cite{bi248,bi257}.

In this work we introduce a different model for the switching, where
the spin-polarization does {\it not relax}. The effect is that
the spin current carries an exchange field that acts on the local
moments of a layer and forces its magnetization to take the
orientation of the spin-polarization of the conduction electrons.

\begin{figure}[tbp]
\begin{center}
\leavevmode
{\hbox {\epsfxsize=3.4in 
\epsffile{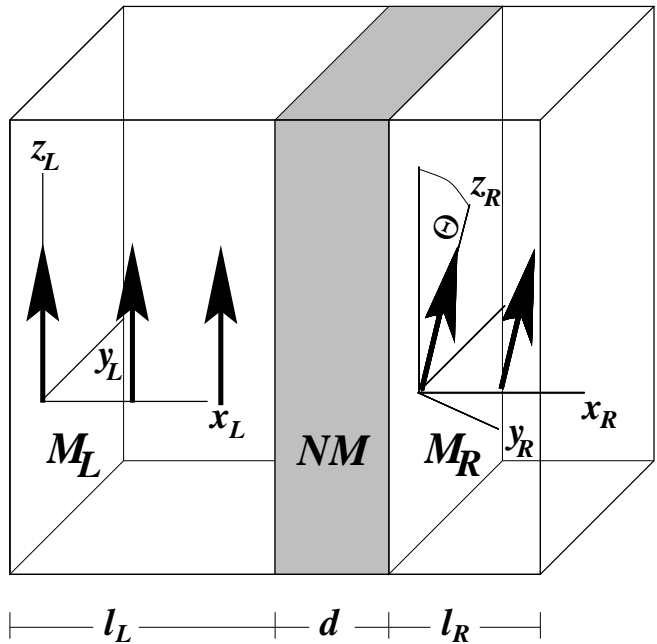}
}} 
\vspace{+3ex} 
\caption{The geometry of a standard trilayer spin-valve; two planar
ferromagnetic metal layers of thickness $l_{L(R)}$ and total magnetic
moments ${\bf M}_{L(R)}$ at an angle $\Theta$ relative to each other
are separated by a non-magnetic metal spacer NM of thickness $d$.}
\label{MIM}
\end{center}
\end{figure}

In the conventional view one considers electrons flowing in the
direction of the net current from, say, a ``fixed'' layer that has a
large magnetic moment to a ``free'' layer that has a small magnetic
moment which is easy to reorient (see Fig.~\ref{MIM}). The effect on
the free layer is indeed relatively small as the associated energy,
which is of the Zeeman-type, is proportional to the product of the
small moment of the free layer and the exchange field of the
conduction electrons polarized by the fixed layer (refer to upper part
of Fig.~\ref{NEXI}).  However, this picture neglects the much larger,
albeit counter intuitive, effect of spin-dependent reflection of
electrons by the free layer, so that spin-polarized electrons move in
the direction opposite to the net current and interact with the large
magnetic moment of the fixed layer. The energy of this contribution is
thus also much larger and of opposite sign (see lower part of
Fig.~\ref{NEXI}).

Taken together, both contributions establish a strong non-equilibrium
exchange interaction (NEXI) between layers~\cite{bi30}. The sign of
this interaction is determined by the direction of the current; it
does not oscillate and its range is controlled by the spin diffusion
length of the conduction electrons~\cite{bi233,anm10}.  Therefore, the
NEXI is a volume effect and exerts a torque (force) throughout
the volume of the magnetic layers (leading to precession); it
is the dominant mechanism that drives the system towards its switching
threshold. Here, we describe a simple model that shows how the
NEXI {\it defines} the magnetic configuration of the layers with
minimum energy, while the relaxation of the conduction electron spin
polarization provides a way for the system to {\it evolve} to its
minimum energy configuration.

The rest of the paper is structured as follows. In Sec.~II we
introduce our model of a spin valve and give a brief derivation of the
NEXI. To illustrate the switching mechanism of the spin valve, in
Sec.~III we derive a set of coupled Landau-Lifshitz equations for
uniform magnetic dynamics, whose structure is similar to those of
antiferromagnets or ferrimagnets; however with the important
difference that they are current driven and the magnetic sublattices
are replaced by the spatially separate magnetic layers.  The following
stability analysis in Sec.~IV shows that depending on the direction of
the current either the parallel or the antiparallel configuration
becomes unstable beyond a critical current, so that, in principle,
even an arbitrarily small amount of relaxation allows a switching of
the magnetically ``softer'', i.e.\ free, layer. Despite the simplicity
of our model, quantitative estimates are in reasonable agreement with
experiments. In Sec.~V we introduce Gilbert damping to describe
possible relaxation processes that allow the system to evolve over
time to its minimum energy configuration, i.e.\ to switch. Finally, we
compare our results to the models by Slonczewski~\cite{bi248,bi257},
and Berger~\cite{bi247} in Sec.~VI and conclude in Sec.~VII.

\section{Non-equilibrium exchange interaction (NEXI)}

The geometry of our model system is a typical spin-valve structure
shown in Fig.~\ref{MIM}. It consists of two planar ferromagnetic metal
layers of thickness $l_{L(R)}$ whose total magnetic moments ${\bf
M}_{L(R)}$ are at an angle $\Theta $ relative to each other. The
magnetic layers are separated by a non-magnetic metallic spacer NM of
thickness $d$.  A complete treatment of the magnetic dynamics of such
a spin valve with a current in the perpendicular direction requires a
simultaneous solution of the equation of motion for individual spins
of the magnetic layers (see for example Ref.~\cite{bi202}), and
equation of motion of the spin-polarized charge carriers~\cite{anm12}.

To focus on the essential effects, the calculations are made on the
simplest assumptions; we model the polycrystalline thin films as
uniformly magnetized layers with uniaxial anisotropy in the plane of
the layers. For clarity, we do not focus on the details that establish
the equilibrium (zero current) coupling between magnetic layers; while
the experiments~\cite{bi253} suggest that the equilibrium
Ruderman-Kittel-Kasuya-Yosida (RKKY) coupling is negligible, the
omission of the dipolar coupling between the layers (fringing fields)
is probably a gross simplification of the physical
picture. We treat the steady state non-equilibrium
(constant current) situation in terms of the NEXI which can be written
as a sum of quantum-interference and current driven terms; see for
example Eqs.~(2) and~(3) in Ref.~\cite{bi233}.  Although the first
contribution to the coupling is finite at equilibrium, {\it a.k.a.}
the RKKY interaction, it is a ``surface'' effect because it oscillates
and scales as $1/d^{2}$~\cite{bi19}, therefore we will neglect it here
as pointed out above. We posit that it is only the current driven
term, which is a volume effect, that is responsible for the switching
of the layers. Its decay is controlled by the spin diffusion length
which is considerably longer than the typical thickness of spacer
layers in metallic multilayers.  Although our calculations are taken
in the ballistic limit, they can be generalized to account for diffuse
transport as will be pointed out in the text.

When a constant current is driven in the direction perpendicular to
the plane of the magnetic layers, it becomes spin polarized.
We define a spin current as: 
\begin{equation}
\label{J0}
\mbox{\boldmath
$\Im$}(j_{L(R)}^{M})=j_{L(R)}^{M}{\bf n}_{L(R)},
\end{equation}
whose polarization is along the unit vector of magnetization ${\bf
n}_{L(R)}= {\bf m}_{L(R)}/m_{L(R)}$ of the local moments, i.e.\ along
the $z_{L(R)}$--axis in Fig.~\ref{MIM}, which {\it generates } the
polarization, and whose magnitude is
\begin{equation}
{j}_{L(R)}^{M}=\frac{\mu _{B}}{e}\left( {j}_{L(R)}^{\uparrow }-{j}
_{L(R)}^{\downarrow }\right)=\frac{\mu _{B}}{e}\,{j}_{{\rm e}}\,\eta
_{L(R)}.  
\label{J1}
\end{equation}
We defined the electric charge and Bohr magneton as $e=-|e|$ and $\mu
_{B}=|e|\hbar /(2m^\ast c)$, respectively; $m^\ast$ being the
effective mass of the conduction electrons. Outside a magnetic layer
the spin current decays within a distance of the spin-diffusion length
$\lambda _{ {\rm sf}}$, so that $\eta _{L(R)}$ is proportional to
$\exp (-x/\lambda _{ {\rm sf}})$ where $x$ is the distance away from
the layer. The factor $\eta _{L(R)}$ describes the spin-dependent
reflection (for diffuse transport this leads to the effect of spin
accumulation and can be included in $\eta _{L(R)}$) and can vary
between -1 and 1.

\begin{figure}[tbp]
\begin{center}
\leavevmode
{\hbox {\epsfxsize=3.4in 
\epsffile{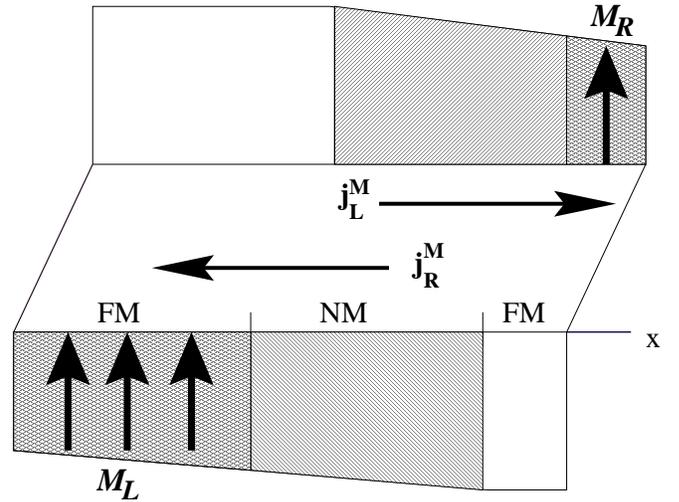}
}} 
\vspace{+3ex} 
\caption{The figure shows the two contributions of 
the spin-current to the non-equilibrium exchange interaction (NEXI) given by 
the hatched regions: the forward contribution $j_L^M$ from the left
ferromagnet and the reflected contribution $j_R^M$ from the right
ferromagnet.  The NEXI between the two
ferromagnetic layers, as shown here for parallel configuration, is
given by the difference of the double hatched regions which are
proportional to $j_L^MM_R$ and $j_R^MM_L$, respectively. Only in
certain limiting cases one of the contributions can be neglected.}
\label{NEXI}
\end{center}
\end{figure}

The NEXI comes then from the coupling of the spin current generated by
the left layer ${\bf \mbox{\boldmath
$\Im$} }(j_{L}^{M})$ interacting with the local
moments in the right layer and vice versa. When we take as the
direction of positive current from left to right, the coupling in
linear response is:
\begin{equation}
E_{{\rm nexi}}^{{\rm eff}}=E_{R}(j_{L}^{M})+E_{L}(-j_{R}^{M}).
\label{EX3}
\end{equation}
The local non-equilibrium coupling $E_{L(R)}$ is proportional to the
scalar product in spin space of the spin current and the local
moments, i.e.\ $ E_{R}\propto {\bf \mbox{\boldmath
$\Im$} }(j_{L}^{M})\cdot {\bf M}_{R}$
and similarly for $E_{L} $, where we averaged the spin-operators over
the non-equilibrium statistical ensemble of the entire system, and
have neglected spin-fluctuations~\cite {bi202}. From Eq.~(\ref{J0}) follows,
${\bf \mbox{\boldmath $\Im$} }(-j_{L(R)}^{M})=-{\bf \mbox{\boldmath
$\Im$} } (j_{L(R)}^{M})$, so
that $E_{L}(-j_{R}^{M})=-E_{L}(j_{R}^{M})$, and Eq.~(\ref{EX3}) takes
the form:
\begin{eqnarray}
E_{{\rm nexi}}^{{\rm eff}} &=&E_{R}(j_{L}^{M})-E_{L}(j_{R}^{M})  
\nonumber \\
&=&\mu _{R}\,{\bf h}_{LR}^{{\rm s-d}}\cdot {\bf M}_{R}-\mu _{L}\,{\bf h}
_{RL}^{{\rm s-d}}\cdot {\bf M}_{L},\label{EX2}
\end{eqnarray}
where ${\bf h}_{RL(LR)}^{{\rm s-d}}=J_{L(R)}{\bf \mbox{\boldmath
$\Im$} }{(j_{R(L)}^{M})}/({ v_{L(R)}^{F}}\mu _{L(R)}\mu _{B})$ are the
local non-equilibrium exchange fields in the mean field
approximation. Here $J_{L(R)}$ is the coupling constant between
conduction electrons and local moments; $\mu _{L(R)}=g_{L(R)}\mu
_{B}$, and $g_{L(R)}$ is the Land\'{e} factor in the respective
layer. The preceding discussion can be visualized by Fig.~\ref{NEXI}.
Another way of writing Eq.~(\ref{EX2}) is in terms of an effective
Heisenberg coupling that depends on the current:
\begin{eqnarray}
E_{{\rm nexi}}^{{\rm eff}} &=&\left(\mu _{R}\,\frac{{h}_{LR}^{{\rm
s-d}}}{M_L}-\mu _{L}\,\frac{{h} _{RL}^{{\rm s-d}}}{M_R}\right)
{\bf M}_{L}\cdot {\bf M}_{R}\nonumber \\
&\equiv&
-J^{{\rm eff}}{\bf M}_{L}\cdot {\bf M}_{R}
.
\label{EX4}
\end{eqnarray}

In Eqs.~(\ref{EX2}) and~(\ref{EX4}), respectively, we also introduced
the simplifying assumption that the amount of spin polarization $\eta
_{L(R)}$ does not depend on the angle $\Theta $ between the magnetic
layers (see Fig.~\ref{MIM}) and that there is no relaxation other than
spin-diffusion. In fact the spin-polarized current of each layer
depends on boundary condition to adjacent layers, e.g., as outlined in
Ref.~\cite{bi27}, so that $\eta _{L(R)} $ has to be calculated
self-consistently from Eqs.~(\ref{J1}) and~(\ref{EX2}), by taking into
account the back effect of the coupling between the layers on the
spin-polarization itself. A simple illustration of an approach which
is self consistent in the current but not in energy can be given in
the limit of diffusive transport by calculating the difference in
energies for parallel and antiparallel configuration associated with
the spin accumulation in the spin-valve. Using the picture developed
in Ref.~\cite{bi27}, one finds that for different layer thickness this
energy contribution is non-vanishing. Details will be given elsewhere
for a fully self-consistent calculation.

The form of Eq.~(\ref{EX2}) leads to some immediate consequences. If
the system is symmetric in its magnetic properties, then, for the
assumed linear response regime of the current, which is reasonably
well satisfied in metallic multilayers even for high current
densities, the two couplings $ E_{L}$ and $E_{R}$ are equal and
opposite, so that there is {\it no} non-equilibrium coupling between
the layers. If, however, the magnetic properties or the thickness of
the magnetic layers are dissimilar (in general, this would include
non-uniform magnetization), the spatial symmetry of the system is
broken and a current driven coupling between the layers exists. In
particular, on reversing the current the interlayer coupling changes
sign.

\section{Equations of motion}

To obtain the equation of motion for the magnetization of the spin
valve, we follow the standard procedure and define an effective
field~\cite{bi258},
\begin{equation}
\label{HE1}
{\bf h}_{L(R)}^{{\rm eff}}={\bf h}_{L(R)}^{(i)}+\beta _{L(R)}{\bf n}_{L(R)}^{
{\rm a}}\left( {\bf m}_{L(R)}\cdot {\bf n}_{L(R)}^{{\rm a}}\right) ,
\end{equation}
which is derived from the total energy of the system including the effects
of uniaxial anisotropy, dipole-dipole interactions, and $E_{{\rm nexi}}^{
{\rm eff}}$ of Eq.~(\ref{EX2}). The internal field (first term in Eqs.~(\ref
{HE1})) includes a contribution from the external field ${\bf h}^{{\rm (e)}}
$,  the spin current, and the magnetic dipole--dipole interaction which in a
uniformly magnetized ellipsoid takes the form 
\begin{mathletters}
\label{HI1}
\begin{eqnarray}
{\bf h}_{L}^{(i)} &=&{\bf h}^{(e)}+{\bf h}_{L}^{\rm nexi}
-4\pi {\bar{N}}
_{L}\cdot {\bf m}_{L},  \label{HI1a} \\
{\bf h}_{R}^{(i)} &=&{\bf h}^{(e)}+{\bf h}_{R}^{\rm nexi}
-4\pi {\bar{N}}
_{R}\cdot {\bf m}_{R},
\end{eqnarray}
\end{mathletters}
where ${\bar{N}}_{L(R)}$ is the tensor of demagnetization factors, and
\begin{mathletters}
\label{HN}
\begin{eqnarray}
{\bf h}_{L}^{\rm nexi} &=&\frac{J^{\rm eff}}{\mu_L}\,{\bf M}_{R}
=\left({h}_{RL}^{{\rm s-d}}
-h_{LR}^{{\rm s-d}}\,\frac{\mu_R}{\mu_L}\frac{M_R}{M_L}\right){\bf n}_R ,
\label{HNL}
\\ 
{\bf h}_{R}^{\rm nexi} &=&\frac{J^{\rm eff}}{\mu_R}\,{\bf M}_{L}
=\left(
h_{RL}^{{\rm s-d}}\,\frac{\mu_L}{\mu_R}\frac{M_L}{M_R}-{h}_{LR}^{{\rm
s-d}}\right){\bf n}_L,
\label{HNR} 
\end{eqnarray}
\end{mathletters}
are the effective non-equilibrium coupling fields between the layers
induced by the spin-current, which vanish for identical layers. In
general, when magnetic layers cannot be described by single domain
particles, this symmetry condition does not hold due to the
non-uniformity of the layers. Further, in such a case the coupling
fields~(\ref{HN}) will contain additional contributions from the
fluctuation of the magnetization. The second term in Eqs.~(\ref{HE1})
is due to the uniaxial anisotropy, where $
\beta _{L(R)}$ is the corresponding constant and ${\bf n}_{L(R)}^{{\rm a}}$
its in-plane unit vector. We would like to point out that the
assumption of uniform magnetization means, in addition, that the
effect of the induced magnetic field from the current, which leads to
vortex formation, is small compared to the coupling field ${\bf
h}^{{\rm nexi}} $ so that we can use the magnetostatic approximation in
which ${\bf h}^{{\rm (e)}}$ is the field produced by currents in coils
alone.  The equations of motion are then given by
\begin{eqnarray}\label{EQM1}
\frac{\partial {\bf M}_{L(R)}}{\partial t} = - \gamma_{L(R)}
\left[ {\bf M}_{L(R)}\times {\bf h}^{\rm eff}_{L(R)} \right] 
+{\bf R}_{L(R)},
\end{eqnarray}
where $\gamma _{L(R)}=g_{L(R)}\mu _{B}/\hbar >0$ are the gyro-magnetic
ratios. The damping terms ${\bf R}_{L(R)}$ will be discussed in detail
later.

Parenthetically, the solution of Eqs.~(\ref{EQM1}) is closely related
to solving the equations of motion for an antiferromagnet or
ferrimagnet, where now an effective field of the exchange coupling
does not act between sublattices but different layers. Therefore, due
to the NEXI there exist long wavelength ``acoustical'' and ``optical''
modes wherein the magnetizations precess either in or out of phase. In
an experimental set-up similar to the one described in
Refs.~\cite{in11,bi251,bi252}, this would offer a direct possibility to
measure the coupling between the magnetic layers as a function of
applied current. The details will be presented elsewhere.

\section{Critical current}

So far we have derived the equations of motion for two coupled
magnetic particles. The solution of the equations requires to divide
the problem into a stationary one, which we shall consider first, and
time-dependent one. In other words, before studying any form of
dynamic behaviour of the magnetization, one must first determine the
steady-state, i.e., the orientations of the magnetizations in the {\it
absence} of time dependent driving fields. This orientation depends on
the current.  In combination with Eqs.~(\ref{EQM1}),
we obtain the equilibrium (steady state or constant current)
configuration of the moments from the conditions:
\begin{equation}
{\bf h}_{L(R)}^{\rm eff}|_{0}\times {\bf M}_{L(R)}=0,  \label{SS1}
\end{equation}
which are independent of the time-dependent damping term and where also
the effective field ${\bf h}_{L(R)}^{\rm eff}|_{0}$ does not depend on
time. These conditions have to be satisfied
simultaneously for both magnetic layers. For many applications, there
exists an interesting limiting case where the thickness and anisotropy
of one of the layers is very large, termed the fixed layer, so that
one can assume, for example, ${\bf n}_{L}= {\bf n}_{L}^{{\rm a}}$. In
other words, the back effect of the right layer, termed the free
layer, will be negligible on the orientation of the left layer but not
on their {\it mutual orientation}. To be more explicit, we study the
steady state of this example in more detail.

Instead of applying the conditions in Eqs.~(\ref{SS1}), a more
straightforward method is to minimize the total energy of the
system. By using the notation we introduced in Fig.~\ref{MIM} and
assuming that the geometry of the film constrains the magnetization to
be in the plane of the layer, the total energy of the system, in the
absence of an external magnetic field, depends only on the angle
$\Theta $ between the magnetic moments of the left and right layer:
\begin{eqnarray}
E(\Theta )=&-&\frac{\beta _{L}}{2}m_{L}M_{L}-\frac{\beta _{R}}{2}
m_{R}M_{R}\cos ^{2}(\Theta -\Theta _{R}^{{\rm a}})  
\nonumber  \\
&-&h_{R}^{{\rm nexi}}M_{R}\cos \Theta ,
 \label{E1}
\end{eqnarray}
where $\Theta _{R}^{{\rm a}}$ is the angle between the easy axis of the
right layer with $z_{L}$, and 
\begin{equation}
h_{R}^{{\rm nexi}}=h_{RL}^{{\rm s-d}}\,\frac{M_{L}}{M_{R}}-h_{LR}^{{\rm s-d}}
\label{HE2}
\end{equation}
is the effective field value of the NEXI on the right layer of
Eq.~(\ref{HNR}) with $\mu_L=\mu_R\equiv \mu$.  From Eq.~(\ref{HE2}) it
follows that for asymmetric magnetic layers a coupling exists which
will be dominated by the spin current generated by the {\it right}
layer if the total magnetic moment of the left layer exceeds that of
the right one, i.e.\ $M_{L}>M_{R}$. In recent experiments one of the
magnetic layers is chosen to be much thicker~\cite{bi253} so that
$M_{L}\gg M_{R}$, and in the following discussion we shall assume this
is the case. From the discussion after Eq.(1) it should be noted that
$h_{LR(RL)}^{{\rm s-d}}$ is proportional to $\lambda _{{\rm
sf}}/l_{R(L)}$ for $l_{L(R)}>\lambda _{{\rm sf}}$ inasmuch as
$j_{L(R)}^{M}$ is limited by the spin diffusion length $\lambda _{{\rm
sf}}$.

It is insightful to express the proportionality between the field
$h_{R}^{{\rm nexi}}$ in Eq.~(\ref{HE2}) and the current density which
generated it by the following relation:
\begin{mathletters}
\label{P1}
\begin{eqnarray}
\Lambda_R^{{\rm nexi}}
 \equiv\frac{ h_{R}^{{\rm nexi}}}{j_e}&=&\frac 1{\mu\, e}\left( 
\frac{J_{L}\eta_{R}}{v_{L}^{F}}
\frac {M_L}{M_R}
-\frac{J_{R}\eta_{L}}{v_{R}^{F}}
\right) 
\label{P1.1} \\
&\approx &\frac {2\,m^\ast}{e}\frac {\eta_L\eta_R}{m_R}\left(
v_{L}^{F}\frac{l_{L}}{l_R}-v_{R}^{F}
\right).  
\label{P1.2} 
\end{eqnarray}
\end{mathletters}
Eq.~(\ref{P1.2}) holds if we assume that the polarizations are the
Pauli factors $\eta _{L(R)}=J_{L(R)}m_{L(R)}/(4\,\varepsilon
_{F}^{L(R)}\mu )$, and $\varepsilon _{F}^{L(R)}$ the Fermi
energies. One can, then, rewrite Eq.~(\ref{HE2}) as:
\begin{equation}
h_{R}^{{\rm nexi}}=\Lambda_{R}^{{\rm nexi}}j_e.
\label{HE3}
\end{equation}
In other words, $h_{R}^{{\rm nexi}}$ plays the role of an exchange
biasing field on the right layer generated by the current through the
system. To get a better understanding, we give a rough estimate to the
magnitude of $\Lambda_R^{{\rm nexi}}$. If we assume a magnetization
for Co of $m_R=1.42\times 10^{6}$ A/m, a Fermi velocity of about
1.5$\times $10$^{6}$ m/s, the layer thickness ratio
$(l_{L}-l_{R})/l_R$=3, and use the free electron mass in
Eq.~(\ref{P1.2}), the only undetermined parameter is $\eta_{L(R)}$.
Taking the hypothetical case that the amount of spin-polarization is
the same throughout the system as in the ferromagnetic layer, we
estimate for Co with $\eta =38$\% $\Lambda_R^{{\rm nexi}}$ to be
approximately 5$\times10^{-5}$ m, which amounts to $1/\Lambda_R^{{\rm
nexi}}\approx 0.02$ mA/kOe.  The latter is about an order of magnitude
lower than the value measured in the experiments on Co/Cu/Co
pillars~\cite{bi267}.  Therefore, our estimate for $\Lambda_R^{{\rm
nexi}}$ derived for perfect spin polarization can be seen as an upper
threshold for the proportionality between biasing field and
current. Realistic estimates are obtained by adjusting the value for
$\eta_{L(R)}$ taking spin-diffusion, spin-dependent reflection at the
interfaces, and the resistance of the layers into
account~\cite{bi30,bi27}. Then for $\eta=10$\%~\cite{bi261}, we obtain
direct agreement with experiments $1/\Lambda_R^{{\rm nexi}}\approx
0.29$ mA/kOe~\cite{bi267}.

Having introduced the effective field $h_{R}^{{\rm nexi}}$ of the
NEXI, we realize from the form of Eq.~(\ref{E1}) that finding the
switching threshold of the right layer (free layer) can be treated as
if the layer were a single magnetic domain in an external magnetic
field. The equilibrium direction of the magnetization of the right
layer (free layer) is then obtained by the extremum energy condition $
(\partial E/\partial \Theta )=0$. In addition, for the equilibrium to
be stable (unstable) the following relation must hold: $(\partial
^{2}E/\partial \Theta ^{2})>(<)$ $0$. At $(\partial ^{2}E/\partial
\Theta ^{2})=0$ the model predicts a transition from a gradual
rotation of ${\bf n} _{R}$ to a sudden switching towards the direction
of ${\bf n}_{L}^{{\rm a}}$, i.e.\ an irreversible magnetization
rotation, which determines the critical coupling field $h_{{\rm
c}}^{{\rm nexi}}$. Since we have two conditions and the two unknown
variables $\Theta $ and $h_{{\rm c}}^{{\rm nexi}}$, we can eliminate
$\Theta $ and derive an expression for the critical coupling field
$h_{{\rm c}}^{ {\rm nexi}}$ obtained from the following relation
(refer to Ref.~\cite{bi268} for the details of the derivation):
\begin{equation}
\sin 2\Theta _{R}^{{\rm a}}=\frac{1}{y^{2}}\left( \frac{4-y^{2}}{3}\right)
^{3/2},  \label{HC1}
\end{equation}
where $y=-2h_{{\rm c}}^{{\rm nexi}}/(\beta _{R}m_{R}^{0})$. From Eq.~(\ref
{HC1}) we find that the critical field is maximum at $\Theta _{R}^{{\rm a}}=0
$ and $\pi /2$, where, $h_{{\rm c}}^{{\rm nexi}}=-\beta _{R}m_{R}^{0}$, and
minimum at $\Theta _{R}^{{\rm a}}=\pi /4$ when $h_{{\rm c}}^{{\rm nexi}
}=-\beta _{R}m_{R}^{0}/2$. For a parallel orientation of the uniaxial
anisotropies between the layers $\Theta _{R}^{{\rm a}}=0$ this translates
together with Eqs.~(\ref{J1}) and~(\ref{HE2}) to the following condition for
a spin current induced switching: 
\begin{mathletters}
\label{IC}
\begin{eqnarray}
-j_{{\rm e}}^{{\rm c}} &\geq&\frac{ 2\,K_R^{\rm u}}{\Lambda_R m_R}
 \label{IC1} \\
&\approx &\frac{e}{m^\ast}
\frac{K_{R}^{{\rm u}}}{\eta _{L}\eta _{R}}\frac{l_{R}}{
l_{L}v_{L}^{F}-l_{R}v_{R}^{F}}  \label{IC2} \\
&\approx &\frac{e}{m^\ast}\frac{K_{R}^{{\rm u}}}{v_{L}^{F}}\frac{1}{\eta
_{L}\eta _{R}}\frac{l_{R}}{l_{L}},  \label{IC3}
\end{eqnarray}
\end{mathletters}
where $j_{{\rm e}}^{{\rm c}}$ is the critical current density, and
$K_{R}^{{\rm u }}=\beta _{R}\,m_{R}^{2}/2$ the standard expression for
the uniaxial anisotropy constant~\cite {bi256}. Eq.~(\ref{IC2}) holds
if we assume again that the polarizations are the Pauli factors, and
Eq.~(\ref{IC3}) is applicable when $M_{L}\gg M_{R}$. From
Eq.~(\ref{IC3}) it becomes clear that the switching threshold is
determined by the coupling of the spin current generated in the right
layer (free layer) {\it to the magnetization in the left layer} (fixed
layer); it would be inappropriate to replace the left layer in the
problem by a simple spin-polarized current source in the switching
experiments~\cite{bi253}. On the contrary, Eqs.~(\ref {IC1})
and~(\ref{IC2}) show that for almost identical layers the critical
current becomes very large as the denominator tends to zero, so that
only transient processes are relevant and, indeed, switching has not been 
observed~\cite{bi259}.  Since $\eta _{L}\eta _{R}$ is a measure of the
coupling strength, the critical current $j_{{\rm e}}^{{\rm c}}$ is
thus inversely proportional to the coupling of the free layer to the
fixed layer, so that a strong coupling reduces $j_{{\rm e}}^{{\rm
c}}$; on the other hand, a high value of anisotropy $K_{R}^{{\rm u}}$
will increase $j_{{\rm e}}^{{\rm c}}$. More generally, $j_{{\rm
e}}^{{\rm c}}$ is proportional not only to the uniaxial anisotropy but
to whatever constrains the local moments; for example $j_{ {\rm
e}}^{{\rm c}}$ will also be strongly influenced by the dipolar
coupling between layers and quantum interference (RKKY) contributions
to the interlayer exchange coupling. Although $K_{R}^{{\rm u}}$
certainly cannot capture the details of the switching, it yields a
simple analytical solution that provides the intuitive picture that
current driven switching can be treated as a single magnetic particle
in the current driven exchange field $ h_{R}^{{\rm nexi}}$. This
description is particularly appealing as it allows one to treat
generalizations of our model similar to those known from magnetic
recording~\cite{bi268}.

We chose $\Theta=0$ to be the parallel alignment of the magnetic
moments for zero current and the direction of current from left to
right which led to a negative sign in front of $h_{R}^{{\rm nexi}}$ in
Eq.~(\ref{E1}). Thus, the parallel configuration is only preferable if
$h_{R}^{{\rm nexi}}$ is positive.  Having related $h_{R}^{{\rm nexi}}$
to the current (refer to Eqs.~(\ref{J1}) and~(\ref{HE2})), the system
becomes then unstable for a sufficient negative current~(\ref{IC})
which forces the spin valve to switch to an antiparallel
configuration. Similarly, starting from an antiparallel orientation, a
positive current switches the spin valve to parallel. We can also give
a rough estimate to the magnitude of required critical current
densities. Using again the same data as before of Ref.~\cite{bi267}
and an uniaxial anisotropy for Co of $K_R^{\rm u}=1.5\times 10^{4}$
J/m$^{3}$, we estimate the lower limit of the critical current for
perfect spin polarization in Eq.~(\ref{IC2}) to be approximately
0.4$\times$10$^{7}$ A/cm$^{2}$; this is more than an order of
magnitude lower than what is needed for the actual switching observed
in the experiments~\cite{bi267}, consistent with our limiting estimate
on $1/\Lambda_R^{{\rm nexi}}$. Using, however, the more realistic
value $\eta=10$\% we obtain a critical current of 0.6$\times$10$^{8}$
A/cm$^{2}$ in good agreement with the experiments~\cite{bi267}.

\section{Relaxation processes}

The conditions for a critical current were obtained from the solution
of the stationary problem. To describe the dynamics of the switching
process, one has to solve the time-dependent problem. In addition, we
have to introduce relaxation, because without it the right layer would
only precess but never switch, no matter how strong the current would
be. However, when the steady state configuration is unstable, even the
slightest relaxation is sufficient for the magnetization to switch. A
wealth of possible relaxation mechanisms arises from the non-uniform
motion of the magnetic moments, which we have excluded here by making
the assumption that the layers can be treated as two coupled uniformly
magnetized particles; also we neglected the back effect of the NEXI on
the spin polarization $\eta _{L(R)}$ which reduces the coupling.  An
important mechanism for relaxation is the transfer of angular momentum
from the conduction electrons to the local moments as discussed by
Slonczewski~\cite{bi248,bi257} and Berger~\cite{bi247}.

Although it is possible to derive this contribution by systematically
taking the perturbation expansion of the $s$--$d$ Hamiltonian to third
order, here we will simply posit it by introducing a phenomenological
Gilbert damping in Eqs.~(\ref{EQM1}):
\begin{equation}
{\bf R}_{L(R)}=\frac{\alpha _{L(R)}}{M_{L(R)}}\;{\bf M}_{L(R)}\times 
\frac{\partial {\bf M}_{L(R)}}{\partial t},  \label{D1}
\end{equation}
where $\alpha _{L(R)}$ is a dimensionless damping parameter that
depends on the current and on the layer thickness. We concentrate only
on the part of $\partial {\bf M}_{L(R)}/\partial t$ due to the NEXI;
all other contributions follow in analogy from the expression of the
effective field~(\ref{HE1}). If we substitute in Eqs.~(\ref{EQM1})
$\gamma _{R}$ with $\gamma _{R}^{\ast }=\gamma _{R}/(1+\alpha _{R})$
we can transform Eq.~(\ref{D1}) into one of the following forms:
\begin{mathletters}
\label{D2}
\begin{eqnarray}
{\bf R}_{R}^{{\rm nexi}} &=&-\frac{\omega _{R}}{M_{R}M_{L}}\;{\bf M}
_{R}\times \left( {\bf M}_{R}\times {\bf M}_{L}\right) 
\label{D2.1} \\
&=&-\omega _{R}\left( {\bf m}_{R}\cos \Theta -\chi _{R}^{0}{\bf
h}_{R}^{ {\rm nexi}}\right) V_{R},
\label{D2.2}
\end{eqnarray}
\end{mathletters}
where $\omega _{R}=\alpha_R \gamma _{R}^{\ast }h_{R}^{{\rm nexi}}$ is
the damping frequency due to the NEXI, $\chi _{R}^{0}=m_{R}/h_{R}^{
{\rm nexi}}$ the respective susceptibility, and $V_{R}$ the volume of
the layer. The first way of writing the relaxation resembles the
effect of the conduction electron spin relaxation of
Refs.~\cite{bi248,bi257}. The result therein, that the surface torques
of both layers impel ${\bf m}_{L}$ and $ {\bf m}_{R}$ to rotate in the
same direction in the plane of the layer, applies also to our case as
can be seen from Eqs.~(\ref{HI1a}) and~(\ref{D1}). However, our
conclusions are different. With reference to Eq.~(\ref{EQM1}), we note
that the NEXI, being the leading order contribution to the switching,
determines the precession so that the effect of ${\bf R} _{L(R)}^{{\rm
nexi}}$ is to reduce the coupling and thus minimize the energy in
Eq.~(\ref{E1}). This is demonstrated by the second way of writing
Eq.~(\ref{D2}) which leads to the interpretation that ${\bf m}_{R}$
relaxes towards the non-equilibrium exchange field ${\bf h}_{R}^{{\rm
nexi}}$ whose orientation is determined by that of the conduction
electron spin polarization which is controlled by ${\bf m}_{L}$ and
the direction of the current.  In other words, if the system becomes
unstable, ${\bf R}_R^{\rm nexi}$ dominates all other contribution in
the Gilbert damping, so that ${\bf R}_R^{\rm nexi}$ drives the system
to its new minimum energy configuration, i.e.\ towards the direction
of ${\bf h}_{R}^{{\rm nexi}}$.  This reasoning is analogous to what
happens to a single magnetic particle in an externally applied
switching field.

Finally, the way in which we introduced the phenomenological damping,
relates the time for the total magnetization to reorient to the
relaxation frequency $\omega_{R}$; the switching motion becomes more
viscous as $\omega_{R}$ becomes large. This coincides with the common
notion that the switching time is expected to be fastest for a
moderate value of $\omega_{R}$, and might allow one to use
Eqs.~(\ref{EQM1}) despite their simplicity to model current driven
switching dynamics in more realistic systems.

\section{Discussion}

We have shown that in order to describe current driven switching of
magnetic layers in sub-micron sized spin-valve structures, it is in
general necessary to include the non-equilibrium coupling between
layers. Only in cases where the system is in linear response and the
magnetic layers are identical, or the spin-diffusion length very
short, one can neglect its contribution.  The assumptions taken in our
model are that of two single domain particles with an uniaxial
anisotropy and a non-equilibrium coupling mediated by the spin-current
and a local $s$--$d$ exchange interaction. These are certainly gross
simplifications of the physical situation, however, they give
quantitative agreement with experiments for a reasonable choice of the
magnetic parameters.

In making comparison with Slonczewski's model~\cite{bi248,bi257}, it
should be noted that we make a different set of assumptions. In his
case the NEXI was neglected and the system was reduced to consisting
of a magnetic layer, that serves as a constant source of spin-polarized
electrons, and a single domain particle which acts as a perfect spin
filter, i.e.\ that all majority spins are transmitted whereas all
minority spins are reflected. The assumption of spin-filtering
distinguishes Slonczewski's theory from that of Berger~\cite{bi247},
who invokes inelastic spin-flip scattering.  However, only inelastic
spin-flip scattering allows for multiple spin-wave excitations that
can reorient the axis of quantization in a magnetic layer and serves
as a different type of switching mechanism in the limit of short
spin-diffusion length, as will be shown further down in the text.

Spin-filtering, which we termed spin-dependent reflection, gives rise
to interlayer coupling. It is non-dissipative to the leading order in
the coupling which is described by the NEXI. Only the next order term
in the interlayer coupling is proportional to Eqs.~(\ref{D2}). By
neglecting the term in leading order, Slonczewski finds only terms
similar to Eq.~(\ref{D2.1}) as the sole origin of multiple spin-wave
excitations as predicted by Berger~\cite{anm14}. Our model produces the
opposite effect: the precession of ${\bf m}_{R}$ is damped and relaxes
towards the non-equilibrium exchange field ${\bf h}_{R}^{\rm
nexi}$. It is also not surprising that after fitting the
phenomenological damping parameter $\alpha_G$~\cite{bi260} and
adjusting $\eta$ to 10\%~\cite{bi261}, which coincides with our
estimate, a good quantitative agreement with experiments is reached;
the competition between the different damping terms ${\bf R}_R$ just
reflects the competition between the different effective field
contributions ${\bf h}^{\rm eff}_R$~(\ref{HE1}) in the Landau-Lifshitz
equation~(\ref{EQM1}).

Nevertheless, Slonczewski's theory is insightful in that it can
describe spin-wave instabilities and shows how switching can occur due
to inelastic spin-flip scattering at a multilayer interface after
adaptation according to Berger~\cite{bi247}. To understand how such a
switching mechanism is in principle possible, one has to consider
first the creation of spin-waves in magnetic multilayers which were
initially observed in point contact measurements of Co/Cu multilayers
above a critical current~\cite{bi259}.  As pointed out before, at the
root of this phenomena lies a relaxation mechanism between conduction
electrons and local moments due to spin-flip scattering. When a
current flows from a non-magnetic metal layer to a ferromagnetic one,
its distribution over spin-up and spin-down currents has to
change. Given that most scattering events will conserve the current in
each spin-subband, a difference in ``chemical potentials'' appears on
the scale of the spin-diffusion length away from the
interfaces~\cite{bi265} which leads to the effect of spin
accumulation~\cite{bi27}. The critical current density for spin-wave
emission is reached when the difference in ``chemical potentials'' of
the spin subbands equals the energy of a spin wave $\Delta\mu
=\mu_{\downarrow}-\mu_{\uparrow }=\hbar \omega$, so that in a
spin-flip transition of a conduction electron the energy is
conserved~\cite{bi247}. In this way magnetic moment is transferred to
the system of localized spins at the non-magnetic/ferromagnetic metal
interface as the conduction electron spin polarization changes.

For spin wave emission a second ferromagnetic layer (e.g.\ the left
layer in Fig.~\ref{MIM}) is not necessarily required as demonstrated
in Fig.~2E of Ref.~\cite{bi253}.  As long as only linear spin-waves
are excited, the transfer of magnetic moment to the local moments
leads just to a reduction of magnetization~\cite{bi254}, similar to
increasing the temperature. The reorientation of the axis of
quantization of the ferromagnet, i.e.\ the switching, takes place {\it
only if} non-linear spin-waves are created~\cite{bi256} which occurs
for much higher current densities. For currents below a critical value
the relaxation mechanism between the conduction electrons and the
local moments allows only for a damping of already existing
spin-waves~\cite{bi247}, similar as in Sec.~V.
 
Since switching seems to occur for current densities well below the
critical value for spin-wave excitations, this mechanism is
insufficient to lie at the origin of the switching. In particular for
Co/Cu/Co spin valves, the data allows us to separate the hysteretic
switching from the spin-wave excitations (refer to Fig.~2 in
Ref.~\cite{bi253}); as mentioned earlier a spin-wave instability is
observed already for an initially unpolarized current. This figure
also shows clearly that the domains are reversed well before the onset
of spin-wave instabilities and at current densities where transport
yields more or less ohmic behaviour.  Thus, experiments seem to
provide evidence that the excitation of spin waves by the spin current
is preceded by the switching process~\cite{bi253,bi267} and,
therefore, confirm our model for the switching based on the NEXI.

\section{Conclusion}

In conclusion, we showed that the current driven coupling provides the
dominant energy term that promotes the switching of the magnetization
of the layers in spin-valves~\cite{bi253,bi267}. We pointed out the
differences to the interpretation that the switching comes about by
creating multiple spin-wave excitations due to inelastic spin-flip
scattering which have to overcome only other forms of relaxation in
the system.

An interesting test of our model is to compare an asymmetric
structure, such as Co/Cu/Py, with a symmetric one, such as
Co/Cu/Py/Cu/Co.  One could also separate the effect of spin-wave
excitations, i.e.\ $\Delta \mu =\hbar \omega $, for large currents
from the switching, if one would apply an external field $h^{(e)}$ to
the spin valve that is much stronger than $h_{R}^{{\rm nexi}}$ given
in Eq.~(\ref{HE2}). On the other hand, interesting effects are
expected to be observed in the region where $h_{R}^{ {\rm nexi}}$ and
$h^{(e)}$ are comparable; in this case much of the magnetic response
of the spin-valve is due to non-uniformity of the magnetization of the
layers which we neglected here.

We are indepted to Peter M.\ Levy for many stimulating discussions and
helpful comments and would like to thank Roger H.\ Koch for the
interest in our work. We are also grateful to John C.\ Slonczewski and
Jonathan Z.\ Sun for communication of results prior to publication.
This work was supported by the Defense Advanced Research Projects
Agency and Office of Naval Research (Grant No.\ N00014-96-1-1207 and
Contract No. MDA972-99-C-0009), and NATO (Grant Ref.\ No.\ PST.CLG
975312). P.E.Z.\ wishes to acknowledge the RFBR (Grant No.\
00-02-16384).

\bibliographystyle{prsty}
\bibliography{/home/carsten/LATEX/biblio/bibs}

\begin{thebibliography}{10}
\bibitem{anm10}
Although in Ref.~\cite{bi30,bi233,bi19} the nonequilibrium exchange interaction
  was calculated for magnetic tunnel junctions, the general results can be
  applied readily to the case of metallic multilayers.
\bibitem{anm12}
This simplified picture uses the notion that transition metals are reasonably
  well described by the $s$--$d$ model; an assumption which holds only well
  within a phenomenological treatment of the coupling.
\bibitem{anm14}
It is interesting to note that in one of the first papers on current induced
  effects in magnetic multilayer systems by Slonczewski~\cite{in17}, the effect
  of the current on what is termed therein conservative exchange coupling was
  not considered.
\end{thebibliography}

\vspace{+5in}

\end{document}